\documentclass[aps, twocolumn, superscriptaddress , amsmath, amssymb, showpacs, showkeys, reprint]{revtex4-1}
\usepackage[utf8]{inputenc}
\usepackage[top = 1.20cm, bottom = 2.10cm, left = 1.6cm, right =1.6cm]{geometry}\usepackage{graphicx}
\usepackage{dcolumn}
\usepackage{bm}
\usepackage{upgreek}
\usepackage{times}
\usepackage{mathrsfs}
\usepackage{mathtools}
\usepackage{multirow}
\usepackage{xcolor}
\usepackage{scalerel}
\usepackage{fancyhdr}
\usepackage{hyperref}
\usepackage{comment}

\newcommand{\nzerocdq}{\Delta n}
\newcommand{\nmodel}{\bar{n}}

\newcommand{\Omegasmall}{\scaleto{\Omega}{3.6pt}}
\newcommand{\charge}{N}
\newcommand{\dipole}{\boldsymbol{\mathcal{D}}}

\newcommand{\centerpos}{\mathbf{R}}
\newcommand{\partialArgRc}{\partial_{\mkern1mu{\centerpos}}}
\newcommand{\partialArgr}{\partial_{\bm{r}}}
\newcommand{\ncells}{\varpi} 
\newcommand{\unknownInt}{\Lambda}
\newcommand{\nosep}{~}

\begin{document}

\title{Model density approach to Ewald summations}
\author{Chiara Ribaldone}
\email[]{chiara.ribaldone@unito.it}
\author{Jacques Kontak Desmarais}
\email[]{jacqueskontak.desmarais@unito.it}
\affiliation{Dipartimento di Chimica, University of Torino, Via Giuria 5, 10125 Torino, Italy}

\begin{abstract}

\centering\begin{minipage}{\dimexpr\paperwidth-7cm}
\noindent
The evaluation of the electrostatic potential is fundamental to the study of condensed phase systems.
We discuss the calculation of the relevant lattice summations by Ewald-type techniques. A model charge density is introduced, that cancels multipole moments of the crystalline charge distribution up to a desired order, for accelerating convergence of the Ewald sums. The method is applicable to calculations of bulk systems, employing arbitrary unit cells in a classical or quantum context, and with arbitrary basis functions to represent the charge density. The efficacy of the method is demonstrated on the calculation of the fundamental gap of the gallium arsenide bulk semiconductor, as a prototype example, where significantly accelerated convergence is numerically confirmed, due to a reduction of the number of two-electron integrals that need to be computed. The approach clarifies a decades-old implementation in the \textsc{Crystal} code. 
\end{minipage}

\end{abstract}

\maketitle

%
%
\section{Introduction}\label{sec:introduction}

Considering a periodic three dimensional system, the series
\begin{equation}\label{sum}
	\sum_{\bm{g}} \frac{1}{\lVert \bm{r} - \bm{g} \rVert}
\end{equation}
where $\bm{g}$ is a lattice vector, is divergent. Eq.\nosep\eqref{sum} is closely related to the problem of computing the electrostatic potential, which must, correspondingly, be handled with care. \\
Under appropriate conditions, this electrostatic problem can be treated successfully. In the class of Evjen-like methods,\cite{evjen_1932, stolarczyk_evjen_like_1982, kudin_evjen_like_1998, kadek_evjen_like_2019, heyes_evjen_like_2026} the lattice sum may be converged by replacing the crystal for a cluster with compensating charges on the surface.
In the class of Ewald-like methods, the sum can be split into two quickly convergent terms, one in direct space and one in reciprocal space, and singularities may be eliminated analytically. Related methods have also been developed using multipole expansions and, in particular, fast multipole methods.\cite{nijboer_ewald_like_1957, challacombe_ewald_like_1997, Aguado_2003, Giese1_2015, burnham_drollery} These Ewald-like methods have enjoyed broad applications in classical\cite{ewald_classical_1989, york_yang_poisson, ewald_classic_1996, ewald_classical_1998, ewald_classical_2001, ewald_classical_2004, zaccone_classical_2019, ewald_classical_2006, bush_2006_daft, ewald_classical_2010, ewald_classical_2015, dl_poly_2026} as well as in quantum mechanical calculations.\cite{ewald_quantum_1996, ewald_quantum_2005, ewald_quantum_2011, ewald_quantum_2012, ewald_quantum_2016, ewald_quantum_2020, ewald_quantum_2022, ewald_quantum_2025} In a recent approach, the crystal is represented as a large finite torus.\cite{tavernier_2020, alrakik_2026} \\
In general, successful direct space bulk lattice summation methods  are based on the fact that the electrostatic potential converges if the charge and dipole, integrated over the unit cell, vanish. In principle, for charge neutral calculations, the dipole can often be nullified by a suitable definition of the shape of the unit cell. However, in practice, the situation is more complicated. For instance, the summation can converge slowly, and the exact shape of the necessary zero dipole cell may vary along a self consistent field calculation. Moreover, the approach breaks down for charged processes, and can be further complicated, in an \emph{ab initio} context, for the calculation of derivatives of the potential. It is therefore desirable to develop extensions of the Ewald procedure whose convergence can be accelerated and rigorously applied to arbitrary unit cells.\\  
An example of such a successful approach is represented by the work of C. Pisani \emph{et al.},\cite{orange_book_pisani} and V. R. Saunders \emph{et al.},\cite{saund_electrostatic} that has enabled a wealth of \emph{ab initio} calculations of materials over the last forty years, through an implementation in the \textsc{Crystal} code. In this approach, a model density is employed to cancel the charge and dipole (as well as higher-order moments), thus leading to quickly convergent electrostatic lattice summations for any unit cell. However, the drawback of the approach lies on its derivation, which invokes rather complicated arguments based on non local {\emph{spreading transformations}} of the electron density, and it is limited in context to \emph{ab initio} calculations that employ Gaussian basis sets to represent the electronic density.\\
In this work, we extend the model density approach to arbitrary basis, and provide a simple derivation, directly established on the form of the electrostatic potential, without relying on spreading transformations. Moreover, a transparent new formula is provided for the representation of the model density.

%
%
\section{Theoretical Method}\label{sec:method}

The theoretical part is organized as follows. In Section \ref{sec:elepot_long}, the convergence problems for the electrostatic potential of an infinite periodic lattice are outlined. In Section \ref{sec:ewald_method}, the Ewald method to treat lattice summations in the conditionally convergent case is presented. In Section \ref{sec:proposing_ewald}, the proposed model density approach to the Ewald summation method is introduced. Finally, Section \ref{sec:nmodel_form} is dedicated to the derivation of a suitable form for the model density.
Before proceeding, note that a glossary of the relevant mathematical quantities is provided in Appendix \ref{sec:glossary}.

\subsection{The electrostatic potential at long range}\label{sec:elepot_long}

Consider a charge distribution $n(\bm{r})$ that has the periodicity of the lattice, i.e.
\begin{equation}\label{ele_long_0}
	n(\bm{r}) = n(\bm{r} \pm \bm{g})
\end{equation}
where $\bm{g}$ is a direct lattice vector.
The form of a lattice component of the electrostatic potential $\Phi[n](\bm{r} - \bm{g})$ generated by $n(\bm{r})$ can be expressed, in Hartree atomic units, as
\begin{equation}\label{ele_long_1}
	\Phi[n](\bm{r} - \bm{g}) = \int_\Omega d^3r' \,\, \frac{n(\bm{r}')}{\lVert \bm{r} - \bm{r}' - \bm{g} \rVert} 
\end{equation}
where $\Omega$ is the unit cell.
As discussed in Appendix \ref{sec:neumann_laplace_derivation}, in the long range (namely, for $\lVert \bm{r} - \bm{g} \rVert > \lVert \bm{r}' \rVert$), the Neumann-Laplace expansion of the inverse distance function leads to
\begin{align}\label{multipole_1}
	& \Phi[n](\bm{r} - \bm{g}) \\
	& \quad = \sum_{\ell = 0}^\infty \,\, \lVert \bm{r} - \bm{g} \rVert^{-(\ell + 1)} \int_\Omega d^3r' \,\, \lVert \bm{r}' \rVert^{\ell} \,\, \mathcal{P}_{\ell}(\cos\gamma) \, n(\bm{r}') \quad \notag
\end{align}
where $\mathcal{P}_{\ell}(\cos\gamma)$ is the Legendre polynomial of degree $\ell$, with $\ell \ge 0$, while $\gamma$ is the angle between $(\bm{r} - \bm{g})$ and $\bm{r}'$ vectors.
Then, writing the Legendre polynomial in terms of spherical harmonics and employing the manipulations of Appendix \ref{sec:neumann_laplace_derivation}, the electrostatic potential of Eq.\nosep\eqref{ele_long_1} at long range can be further rewritten as
\begin{align}\label{ele_long_7}
\begin{aligned}
	& \Phi[n](\bm{r} - \bm{g}) = \int_\Omega d^3r' \,\, \frac{n(\bm{r}')}{\lVert \bm{r} - \bm{r}' - \bm{g} \rVert} \\
	& \quad = \sum_{\ell = 0}^\infty \sum_{m = -\ell}^{\ell} \eta_{\ell\Omegasmall}^m[n](\centerpos) \, Z_\ell^m(\partialArgRc) \, \lVert \bm{r} - \centerpos - \bm{g} \rVert^{-1}
\end{aligned}
\end{align}
where $Z_\ell^m(\partialArgRc)$ is the real spherical gradient operator, defined in Eqs.\nosep\eqref{gloss_5} and \eqref{gloss_6}, with quantum numbers $\ell$ and $m$ (with $-\ell \le m \le \ell$), while $\centerpos$ is an atomic position defining the center of the expansion, and $\eta_{\ell\Omegasmall}^m[n](\centerpos)$ represents the cell multipole moments of the charge distribution $n(\bm{r})$, defined as 
\begin{equation}\label{ele_long_8}
	\eta_{\ell\Omegasmall}^m[n](\centerpos) = \int_\Omega d^3r \,\, n(\bm{r}) \, X_\ell^m(\bm{r} - \centerpos)
\end{equation}
where $X_\ell^m(\bm{r} - \centerpos)$ are the unnormalized real solid spherical harmonics, as given by Eqs.\nosep\eqref{gloss_3} and \eqref{gloss_4}.

\subsubsection{Convergence of the lattice series}\label{sec:integral_test}

Consider a lattice with $\ncells$ periodically repeated cells, each of which is associated to a direct lattice vector $\bm{g}$. The corresponding electrostatic potential at a generic point $\bm{r}$ can thus be evaluated by writing the expansion in Eq.\nosep\eqref{multipole_1} in a compact form and summing the contributions of the resulting expression over the whole set of direct lattice vectors, thus leading to
\begin{equation}\label{cauchy_test_1}
\begin{aligned}
    \Phi[n](\bm{r}) & = \sum_{\bm{g}} \Phi[n](\bm{r} - \bm{g}) = \sum_{\ell = 0}^\infty \sum_{\bm{g}} \frac{c_\ell[\hat{\bm{u}}(\bm{r} - \bm{g})]}{\lVert \bm{r} - \bm{g} \rVert^{\ell+1}}
\end{aligned}
\end{equation}
where the unit vector $\hat{\bm{u}}(\bm{r} - \bm{g}) = (\bm{r} - \bm{g})/\lVert \bm{r} - \bm{g} \rVert$ has been defined, and the coefficients $c_\ell$ depend only on the direction of $(\bm{r} - \bm{g})$ and not on its length.
Inspection of the convergence properties of Eq.\nosep\eqref{cauchy_test_1} through the Maclaurin-Cauchy test\cite{knopp_1958,whittaker_1996} shows that the convergence of the lattice series depends on the convergence of the following improper integral
\begin{equation}\label{cauchy_test_2}
	\mathcal{F}_{\ell d} = \int_1^\infty dh \,\, h^{d \, - \, \ell \, - \, 2}
\end{equation}
in $d$ spatial dimensions.
In particular, for $d = 3$, the absolute convergence of the lattice series requires a vanishing charge ($\ell=0$), dipole ($\ell=1$) and quadrupole ($\ell=2$). On the other hand, if the quadrupole is finite (but the charge and dipole vanish), Eq.\nosep\eqref{cauchy_test_2} acquires a logarithmic singularity, and thus the series is rendered conditionally convergent.
This problem was studied by P. P. Ewald and others,\cite{ewald_1921, leeuw_1980, makov_1995} who developed summation techniques for the conditionally convergent case, as discussed in the following section.

\subsection{The Ewald electrostatic potential}\label{sec:ewald_method}

The method consists in the introduction, in the expression for the electrostatic potential
\begin{equation}\label{ewald_0}
	\Phi[n](\bm{r}) = \sum_{\bm{g}} \int_\Omega d^3r' \,\, \frac{n(\bm{r}')}{\lVert \bm{r} - \bm{r}' - \bm{g} \rVert} 
\end{equation}
of a Gaussian convergence factor, depending on a parameter $s \ge 0$ in the following way
\begin{equation}\label{ewald_1}
	\Phi[n](\bm{r},s) = \sum_{\bm{g}} \int_{\Omega} \, d^3r' \,\, \frac{n(\bm{r}')}{\lVert \bm{r} - \bm{r}' - \bm{g} \rVert} \,\, e^{-s\lVert \bm{g} \rVert^2}
\end{equation}
so that, in the limit $s \to 0$, the electrostatic potential of Eq.\nosep\eqref{ewald_1} recovers the exact form of Eq.\nosep\eqref{ewald_0} \textit{if the lattice series in Eq.\nosep\eqref{ewald_0} converges}. Let us now consider the conditionally convergent case in three dimensions by introducing a generic difference of charge densities $\nzerocdq(\bm{r})$, defined such that the associated charge of the unit cell, as well as all the components of the unit cell dipole moment vector, are zero, that is
\begin{equation}\label{ndiff}
	\charge_{\Omegasmall}[\nzerocdq] = \mathcal{D}_{i \Omegasmall}[\nzerocdq] = 0 \qquad \quad \forall \,\, i
\end{equation}
It can be demonstrated (see Appendix \ref{sec:ewald_derivation}) that the limit $s \to 0$ of Eq.\nosep\eqref{ewald_1} leads to an electrostatic potential of the form 
\begin{align}\label{ewald_9}
	\Phi[\nzerocdq](\bm{r}) & = \Phi^{\text{ew}}[\nzerocdq](\bm{r}) \\
	& = \int_{\Omega} d^3r' \,\, \nzerocdq(\bm{r}') \, A(\bm{r} - \bm{r}',\kappa) \, - \, \frac{2\pi}{3v} \, \mathcal{S}_{\Omegasmall}[\nzerocdq] \notag
\end{align}
where $v$ is the volume of the unit cell, $\mathcal{S}_{\Omegasmall}[\nzerocdq]$ is the so called spheropole, representing a shape-dependent shift in the potential and given by 
\begin{equation}\label{cell_spheropole}
    \mathcal{S}_{\Omegasmall}[n] = \int_{\Omega} d^3r \,\, n(\bm{r}) \, \lVert \bm{r} \Vert^2
\end{equation}
and a new quantity has been introduced, as
\begin{equation}\label{ewald_13}
	A(\bm{r} - \bm{r}',\kappa) \, = \, \Xi(\bm{r} - \bm{r}',\kappa) \, - \, \frac{\pi}{v \kappa}
\end{equation}
where $\kappa$ is a real number ($\kappa > 0$) and the function on the right hand side is given by
\begin{equation}\label{ewald_40}
\begin{aligned}
	\Xi(\bm{r} - \bm{r}',\kappa) & = \, \sum_{\bm{g}} \frac{\text{erfc}\left(\sqrt{\kappa} \, \lVert \bm{r} - \bm{r}' - \bm{g} \rVert\right)}{\lVert \bm{r} - \bm{r}' - \bm{g} \rVert} \\
	& + \, \frac{4\pi}{v} \sum_{\mathbf{G} \neq \bm{0}} \left( \frac{e^{-\lVert\mathbf{G}\rVert^2/(4\kappa)}}{\lVert \mathbf{G} \rVert^2} \right) e^{i\mathbf{G}\cdot(\bm{r} \, - \, \bm{r}')} \,\,\,
\end{aligned}
\end{equation}
where $\mathbf{G}$ denotes a reciprocal lattice vector and the parameter $\kappa$ controls the relative extent of direct and reciprocal lattice summations in the function $\Xi(\bm{r} - \bm{r}',\kappa)$. Specifically, small $\kappa$ values leads to a rapid convergence in reciprocal space, but a slowly convergent series in direct space, and vice versa for large $\kappa$. 
Moreover, the contribution to the potential from the constant term $-\pi/(v\kappa)$ appearing in Eqs.\nosep\eqref{ewald_9} and \eqref{ewald_13} vanishes, under the condition of charge neutrality, if the same value of $\kappa$ is used for electrons and nuclei, but may be non-vanishing if different values are employed.\cite{saund_electrostatic} \\
It is worth emphasizing that the identity in the first line of Eq.\nosep\eqref{ewald_9}, which equates the exact electrostatic potential of Eq.\nosep\eqref{ewald_0} with its Ewald representation, is valid only when the charge density involved fulfills Eq.\nosep\eqref{ndiff}.
In practice, a charge density difference $\Delta n(\bm{r})$ that guarantees vanishing quadrupole, that is
\begin{equation}\label{ewald_46}
\mathcal{Q}_{ij \Omegasmall}[\nzerocdq] = 0 \qquad \quad \forall \,\, i,j
\end{equation}
can be considered, to ensure absolute convergence. Extending this same condition to higher order moments, the absolute convergence can be achieved faster. \\
In Eq.\nosep\eqref{ewald_9}, all the integrals are taken over the unit cell $\Omega$. However, for analytical evaluations, it is convenient to extend the integration to the entire space. 
This can be achieved by noticing that the function $A(\bm{r} - \bm{r}', \kappa)$ in Eqs.\nosep\eqref{ewald_13} and \eqref{ewald_40} is lattice periodic, and that Eq.\nosep\eqref{ndiff} guarantees the lattice periodicity of the spheropole $\mathcal{S}_{\Omegasmall}[\nzerocdq]$ in Eq.\nosep\eqref{ewald_9}, thus leading to
\begin{align}\label{int_ext_4}
    & \Phi[\nzerocdq](\bm{r}) = \Phi^{\text{ew}}[\nzerocdq](\bm{r}) \\
    & \quad = \frac{1}{\ncells} \int_{-\infty}^\infty d^3r' \,\, \nzerocdq(\bm{r}') \, A(\bm{r} - \bm{r}', \kappa) \, - \, \frac{2\pi}{3v\ncells} \, \mathcal{S}[\nzerocdq] \quad \notag
\end{align}
where the spheropole term is given by
\begin{equation}\label{spheropole}
	\mathcal{S}[n] = \int_{-\infty}^\infty d^3r \,\, n(\bm{r}) \, \lVert \bm{r} \Vert^2
\end{equation}
Note that $\mathcal{S}[\nzerocdq]$ of Eq.\nosep\eqref{int_ext_4} merely represents a  shift of the electrostatic potential. In the case of charge neutrality of the unit cell, this shift is not relevant for the calculation of the total energy of the crystal, but may be involved in the computation of energy derivatives in a finite basis, due to the \emph{Pulay force} contributions.\cite{saund_electrostatic, jacques_grad, maschio_raman, maschio_ir, doll_grad}

\subsection{Model density approach to Ewald summation}\label{sec:proposing_ewald}

Applying Eqs.\nosep\eqref{ndiff} and \eqref{ewald_46} to each cell in the system, the three conditions can be extended to the entire space
\begin{equation}\label{ndiff_entire_space}
\charge[\nzerocdq] = \mathcal{D}_{i}[\nzerocdq] = \mathcal{Q}_{ij}[\nzerocdq] = 0 \qquad \quad \forall \,\, i,j
\end{equation}
and may be fulfilled by introducing a charge density difference $\nzerocdq(\bm{r})$ with the form
\begin{equation}\label{ndiff_1}
    \nzerocdq(\bm{r}) = n(\bm{r}) - \nmodel(\bm{r})
\end{equation}
where $\nmodel(\bm{r})$ is a model density, defined to reproduce the monopole, dipole and quadrupole moments of the exact total charge distribution $n(\bm{r})$. 
In view of Eq.\nosep\eqref{ndiff_1} for the charge density difference, Eq.\nosep\eqref{int_ext_4} can be rewritten as
\begin{equation}\label{ewald_prop_1}
\begin{aligned}
& \Phi[\nzerocdq](\bm{r}) = \frac{1}{\ncells} \int_{-\infty}^\infty d^3r' \,\, n(\bm{r}') \, A(\bm{r} - \bm{r}', \kappa) \\
& \quad - \frac{1}{\ncells} \int_{-\infty}^\infty d^3r' \,\, \nmodel(\bm{r}') \, A(\bm{r} - \bm{r}', \kappa) \, - \, \frac{2\pi}{3v\ncells} \, \mathcal{S}[\nzerocdq]
\end{aligned}
\end{equation}

Rearranging the terms in the previous Eq.\nosep\eqref{ewald_prop_1} leads to the expression
\begin{equation}\label{ewald_prop_2}
\begin{aligned}
 & \frac{1}{\ncells} \int_{-\infty}^\infty d^3r' \,\, n(\bm{r}') \, A(\bm{r} - \bm{r}', \kappa) \\
 & \qquad \qquad = \, \frac{1}{\ncells} \int_{-\infty}^\infty d^3r' \,\, \nmodel(\bm{r}') \, A(\bm{r} - \bm{r}', \kappa) \quad \\
 & \qquad \qquad \quad \,\, + \, \Phi[\nzerocdq](\bm{r}) \, + \, \frac{2\pi}{3v\ncells} \, \mathcal{S}[\nzerocdq]
\end{aligned}
\end{equation}
Eq.\nosep\eqref{ewald_prop_2} is our main result. It allows the straightforward evaluation of the electrostatic potential in an absolutely convergent manner, provided any suitable choice of the model density $\nmodel(\bm{r})$ and for arbitrary unit cells. Eq.\nosep\eqref{ewald_prop_2} is similar to equation (116a) of Ref.\nosep\cite{saund_electrostatic}, but with the total charge distribution $n(\bm{r})$ instead of a lattice component. Notably, we were able to derive Eq.\nosep\eqref{ewald_prop_2} directly from the form of the electrostatic potential, without invoking more involved arguments based on non local spreading transformations, and without introducing assumptions on the basis functions employed to represent the charge density. In the following section, an explicit form of the model density $\nmodel(\bm{r})$ is provided.

\subsection{The form of the model density}\label{sec:nmodel_form}

Eqs.\nosep\eqref{ndiff_entire_space} and \eqref{ndiff_1} can be generalized to arbitrary multipole moments as
\begin{equation}\label{nmodel_3}
 \eta_\ell^m[\nmodel](\centerpos) = \eta_\ell^m[n](\centerpos)
\end{equation}
where the expression for the multipole moments is given by
\begin{equation}\label{nmodel_2}
    \eta_\ell^m[n](\centerpos) = \int_{-\infty}^\infty d^3r \,\, n(\bm{r}) \, X_\ell^m(\bm{r} - \centerpos)
\end{equation}

To satisfy Eq.\nosep\eqref{nmodel_3} up to an arbitrary order $\ell = L$ (where, in particular, the three requirements in Eq.\nosep\eqref{ndiff_entire_space} correspond to the case $L = 2$), the model density can be expanded in terms of multipole moments of $n(\bm{r})$ around the point $\mathbf{R}$, as
\begin{equation}\label{nmodel_4}
    \nmodel(\bm{r}) = \sum_{\ell = 0}^L \sum_{m = -\ell}^{\ell} \eta_\ell^m[n](\centerpos) \, \varkappa^m_{\ell}(\bm{r} - \centerpos)
\end{equation}
where the explicit form of the model function $\varkappa^m_{\ell}(\bm{r} - \centerpos)$ is to be defined through the requirement that Eq.\nosep\eqref{nmodel_3} is satisfied. In practice, a different center for the multipole expansion $\centerpos$ can be employed, for instance, for different basis functions.\cite{orange_book_pisani} Inserting Eq.\nosep\eqref{nmodel_4} into Eq.\nosep\eqref{nmodel_2} for the model density leads to
\begin{align}\label{nmodel_5}
\begin{aligned}
    \eta_\ell^m[\nmodel](\centerpos) & = \int_{-\infty}^\infty d^3r \,\, \nmodel(\bm{r}) \, X_\ell^m(\bm{r} - \centerpos) \\
    & = \sum_{\ell' = 0}^L \sum_{m' = -\ell'}^{\ell'} \eta_{\ell'}^{m'}[n](\centerpos) \, \unknownInt_{\ell\ell'}^{m m'}
\end{aligned}
\end{align}
where the unknown function
\begin{equation}\label{nmodel_5_1}
	\unknownInt_{\ell\ell'}^{m m'} = \int_{-\infty}^\infty d^3r \,\, \varkappa^{m'}_{\ell'}(\bm{r} - \centerpos) \, X_\ell^m(\bm{r} - \centerpos)
\end{equation}
has been introduced. Comparing Eq.\nosep\eqref{nmodel_5} to Eq.\nosep\eqref{nmodel_3}, the requirement that the model density reproduces an arbitrary moment of the exact density leads immediately to the condition
\begin{equation}\label{nmodel_6}
    \unknownInt_{\ell\ell'}^{m m'} = 
    \delta_{\ell\ell'} \, \delta_{mm'}
\end{equation}

A form of the model function which satisfies Eq.\nosep\eqref{nmodel_6} can be written as
\begin{equation}\label{nmodel_7}
\begin{aligned}
    \varkappa^{m}_{\ell}(\bm{r} - \centerpos) = \mathcal{R}_\ell^m(\lVert \bm{r} - \centerpos \rVert) \, X_\ell^m(\bm{r} - \centerpos)
\end{aligned}
\end{equation}
where $\mathcal{R}_\ell^m$ is an unknown radial function.
Inserting the form of Eq.\nosep\eqref{nmodel_7} inside Eq.\nosep\eqref{nmodel_5_1}, using the translational invariance in the integration over the whole space and then changing from Cartesian to spherical variables, provides
\begin{equation}\label{nmodel_9}
\begin{aligned}
	& \unknownInt_{\ell\ell'}^{m m'} = \mathcal{A}_{\ell \ell'}^{m m'} \int_0^\infty dr \,\, \mathcal{R}_{\ell'}^{m'}(r) \,\, r^{\ell \, + \, \ell' \, + \, 2}
\end{aligned}
\end{equation}
where the angular integral $\mathcal{A}_{\ell \ell'}^{m m'}$ can be solved, as detailed in Appendix \ref{sec:angular_integral}, using Eq.\nosep\eqref{gloss_3} for the unnormalized real spherical harmonics, together with the Theorem 3.11 of Ref.\nosep\cite{bell_1968}, thus resulting in the simple expression
\begin{equation}\label{nmodel_10}
\begin{aligned}
	\mathcal{A}_{\ell \ell'}^{m m'} & = \int_0^{2\pi} d\varphi \int_0^\pi d\theta \,\, X_\ell^m(\theta, \varphi) \, X_{\ell'}^{m'}(\theta, \varphi) \, \sin\theta \\\
	& = \left(C_\ell^m\right)^{-1} \, \delta_{\ell\ell'} \, \delta_{mm'}
\end{aligned}
\end{equation}
where the constant factors have been collected in the quantity
\begin{equation}\label{nmodel_11}
	C_\ell^m =  \frac{(2\ell + 1)(2 - \delta_{m0})(\ell - \lvert m \rvert)!}{4\pi(\ell + \lvert m \rvert)!}
\end{equation} 
Substituting Eq.\nosep\eqref{nmodel_10} in Eq.\nosep\eqref{nmodel_9} gives
\begin{align}\label{nmodel_12}
	& \unknownInt_{\ell\ell'}^{m m'} = \int_{-\infty}^\infty d^3r \,\, \varkappa^{m'}_{\ell'}(\bm{r} - \centerpos) \, X_\ell^m(\bm{r} - \centerpos) \\
	& \quad = \left(C_\ell^m\right)^{-1} \, \delta_{\ell\ell'} \, \delta_{mm'} \int_0^\infty dr \,\, \mathcal{R}_{\ell}^{m}(r) \,\, r^{2(\ell \, + \, 1)} \notag
\end{align}
and, thus, using the definition of the \emph{strong} radial Dirac delta function (see Eq.\nosep(1.1.5) of Ref.\nosep\cite{barton_green_functions_1989}), that is
\begin{equation}\label{strong_dirac}
    \int_0^\infty dr \, \delta(r) \, f(r) = f(0) 
\end{equation}
then Eqs.\nosep\eqref{nmodel_6} and \eqref{strong_dirac} imply
\begin{equation}\label{nmodel_13}
\begin{aligned}
	 \mathcal{R}_\ell^m(r) = C_\ell^m \, \delta(r) \, r^{-2(\ell + 1)}
\end{aligned}
\end{equation}
which, upon insertion into Eq.\nosep\eqref{nmodel_7} and then in Eq.\nosep\eqref{nmodel_4}, provides a simple form for the model density.
This, in turn, leads to simple calculations for the integrals in Eq.\nosep\eqref{ewald_prop_2}, through the use of Eq.\nosep\eqref{ele_long_5}, for instance
\begin{equation}
\begin{aligned}
    & \int_{-\infty}^\infty d^3r' \,\, \nmodel(\bm{r}') \, A(\bm{r} - \bm{r}', \kappa) \\
    & \qquad = \sum_{\ell = 0}^L \sum_{m = -\ell}^{\ell} \eta_{\ell}^m[n](\centerpos) \, Z_\ell^m(\partialArgRc) A(\bm{r} - \mathbf{R}, \kappa)
\end{aligned}
\end{equation}

as well as

\begin{equation}
\begin{aligned}
    & \int_{-\infty}^\infty d^3r' \,\, \nmodel(\bm{r}') \, \lVert \bm{r} - \bm{r}' \rVert^{-1} \\
    & \qquad = \sum_{\ell = 0}^L \sum_{m = -\ell}^{\ell} \eta_{\ell}^m[n](\centerpos) \, Z_\ell^m(\partialArgRc) \lVert \bm{r} - \mathbf{R} \rVert^{-1}
\end{aligned}
\end{equation}
The analytical expression for the spheropole integral $\mathcal{S}[\nmodel]$ can also be readily computed.  

%
%

\section{Numerical example}\label{sec:numerical_results}
As a representative numerical example of the efficiency of the model density method, we consider the GaAs crystal, a direct-gap semiconductor that crystallizes in the cubic $\bar{F}$43$m$ space group (i.e. the zincblende structure). \\
Using the \textsc{Crystal} code,\cite{crystal23} we employed the Gaussian basis set of Ref.\nosep\cite{heydt_2005}, a dense Monkhorst-Pack net of $60\times60\times60 \ \bm{k}$ points, and the Perdew Burke Ernzerhof generalized gradient approximation ({\small GGA}) to the exchange-correlation functional of density functional theory.\cite{pbe} \\
The geometry was fully optimized and the computed fundamental gaps are reported in Table\nosep\ref{tab:results}, as a function of the maximum index $N_{\bm{g}}$ of the vector included in the lattice summations, and of the maximum multipole order $L$ used to represent the model density in Eq.\nosep\eqref{nmodel_4}. In the implementation, the number of terms $N_\mathbf{g}$ for the lattice summations is selected based on a cutoff tolerance on the overlap integral of the Gaussian basis functions (i.e. ``Tol.'' provided in the first column of Table\nosep\ref{tab:results}). An increased $N_{\bm{g}}$ leads to an increased cost of the calculation, principally through the two-electron Hartree integrals that need to be evaluated for the energy contribution 
\begin{equation}
    E[n, \Delta n] = \int_{-\infty}^\infty d^3 r \,\, n(\bm{r}) \, \Phi[ \Delta n](\bm{r})
\end{equation}
due to the second term on the right hand side of Eq.\nosep\eqref{ewald_prop_2}. Considering that each two-electron integral has three lattice vector indices, and without invoking possible approximations, the number of integrals that need to be evaluated in a calculation scales as $N_I = N_{\bm{g}}^3$ (third column of Table\nosep\ref{tab:results}).

\begin{table*}[ht!]
    \renewcommand{\arraystretch}{1.4}
    \setlength{\tabcolsep}{10pt}
    \centering
    \begin{tabular}{|lcc|ccccc|}
         \multicolumn{1}{c}{} & \multicolumn{1}{c}{} &
         \multicolumn{1}{c}{} &
         \multicolumn{5}{c}{$E_g$ [eV]} \\
         \cline{4-8}
         \multicolumn{1}{c}{Tol.} & $N_{\bm{g}}$ & $N_I$ & $L=2$ & $L=3$ & $L=4$ & $L=5$ & $L=6$ \\
         \hline
         $10^{-6}$ & 81 & 5.314$\,\cdot\,10^5$& 0.2351 & 0.1814 & 0.1922 & 0.1922 & 0.1968 \\
         $10^{-12}$ & 207 & 8.870$\,\cdot\,10^6$ & 0.1259 & 0.1952 & 0.1954 & 0.1954 & 0.1969 \\
         $10^{-20}$ & 387 & 5.796$\,\cdot\,10^7$ & 0.1990 & 0.1979 & 0.1966 & 0.1966 & 0.1966 \\
         $10^{-24}$ & 477 & 1.085$\,\cdot\,10^8$ & 0.1912 & 0.1954 & 0.1967 & 0.1967 & 0.1967 \\
         $10^{-30}$ & 677 & 3.103$\,\cdot\,10^8$ & 0.2061 & 0.1962 & 0.1967 & 0.1967 & 0.1967 \\
         $10^{-36}$ & 899 & 7.266$\,\cdot\,10^8$& 0.1928 & 0.1965 & 0.1966 & 0.1966 & 0.1967 \\
         $10^{-42}$ & 1067 & 1.215$\,\cdot\,10^9$& 0.1945 & 0.1965 & 0.1966 & 0.1966 & 0.1967 \\
         $10^{-48}$ & 1301 & 2.202$\,\cdot\,10^9$ & 0.2041 & 0.1965 & 0.1967 & 0.1967 & 0.1967 \\
         \hline
    \end{tabular}
    \caption{Fundamental gap $E_g$ [in eV] for the GaAs semiconductor (with the generalized gradient approximation) as a function of the maximum index $N_{\bm{g}}$ of vectors included in the lattice sums, as well as maximum multipole order $L$ to define the model density $\bar{n}(\bm{r})$ through equation \eqref{nmodel_4}. $N_I = N_{\bm{g}}^3$ is proportional to the number of two-electron integrals that need to be correspondingly evaluated in the calculation.}\label{tab:results}
\end{table*}

Regarding the multipole order $L$ of the model density, formally, $L=1$ ensures conditional convergence of the lattice series, and $L=2$ ensures absolute convergence. Then, values $L>2$ are desirable to accelerate the convergence of the lattice series. This behaviour can indeed be verified in Table\nosep\ref{tab:results}. For small values of $L$ (e.g. $L=2$), we are unable to obtain a highly precise value of the gap (0.197 eV), even by including $N_{\bm{g}}=1301$ terms in the lattice series. On the other hand, for $L=6$, we get a converged value of the gap already with just $N_{\bm{g}}=81$ terms  in the lattice summation (and intermediate results are reported for $L=3,4,5$), which represents a significant acceleration for the convergence of the series (saving a factor of over one order of magnitude on $N_{\bm{g}}$ and over three orders of magnitude on the number of two-electron integrals). These same savings are also reflected in the computational timings, reported in Table\nosep\ref{tab:results_timings} of Appendix\nosep\ref{sec:computational_timings}, where a gain of over three orders of magnitude on the timings can be observed.

%
%
\section{Conclusions}\label{sec:conclusions}

In this work, the model density approach to Ewald summations has been extended to arbitrary basis functions representing the crystalline charge density. The form of the model density was found directly from the requirement that the electrostatic potential converges, instead of relying on more complicated arguments based on spreading transformations of the charge density. Having established the transparency of the approach, it is tantalizing to extend the method to future developments, that improve computational efficiency and enable \emph{ab initio} calculations for complex properties of materials. In the long term, we look forward to apply the approach for the computation of the orbital Hessian in periodic systems, for the time-dependent density functional theory and for employing more advanced density functionals.

%
%
\begin{acknowledgments}
This research has received funding from the project {\small CH4.0}, sponsored by the Italian institution Ministero dell'Università e della Ricerca ({\small MUR}), under the program “Dipartimenti di Eccellenza {\small 2023-2027}” ({\small CUP: D13C22003520001}). Chiara Ribaldone also acknowledges the financial support from {\small PRIN} project no. {\small 2022LM2K5X}, Adiabatic Connection for Correlation in Crystals ({\small AC3}) funded by European Union - NextGeneration{\small EU} - {\small PNRR}. We thank Dr. Stefano Pittalis as well as Dr. Ian J. Bush for suggestions which have helped improving the manuscript.
\end{acknowledgments}


%
%

\appendix

\section{Glossary}\label{sec:glossary}

\subsection{Real spherical harmonics and Legendre polynomials}

The unnormalized real spherical harmonics $X_\ell^m(\theta, \varphi)$ are defined as\cite{orange_book_pisani}
\begin{equation}\label{gloss_3}
	\begin{aligned}
		& X_\ell^m(\theta, \varphi) = \mathcal{P}_\ell^{\lvert m \rvert}(\cos\theta) \, t_m(\varphi) \\
		& \text{where} \qquad 
		t_m(\varphi) = 
		\begin{cases}
			\cos(m \varphi) & \text{for} \quad m \ge 0 \\
			\sin(-m \varphi) & \text{for} \quad m < 0 \\
		\end{cases}
	\end{aligned}
\end{equation}
where $\mathcal{P}_\ell^{m}(\cos\theta)$ are the unnormalized associated Legendre polynomials, defined as in equation (12.144) of Ref.\nosep\cite{math_phys_arfken}.
Eq.\nosep\eqref{gloss_3} represents the angular part of the unnormalized real solid spherical harmonics, given by
\begin{equation}\label{gloss_4}
	X^m_\ell(\bm{r}) = \lVert \bm{r} \rVert^\ell \, X^m_\ell(\theta, \varphi)
\end{equation} 

In Cartesian coordinates, the real spherical gradient operator is defined as 
\begin{equation}\label{gloss_5}
	X_\ell^m(\partialArgr) = \sum_{tuv} \mathcal{D}_{tuv}^{\ell m} \,\, \partial_x^t \, \partial_y^u \, \partial_z^v
\end{equation}
where $t + u + v = \ell \ge 0$ and the shortest notation $\partial_\mu \equiv \partial / \partial r_\mu$ with $\mu = x, y, z$ and $\bm{r} = (r_x, r_y, r_z)$ has been used. \\
The expansion coefficients in Eq.\nosep\eqref{gloss_5} can be computed using the formula derived and discussed in Ref.\nosep\cite{spherical_2025}.
The normalized form of the real spherical gradient operator reads\cite{saund_electrostatic}
\begin{equation}\label{gloss_6}
	Z_\ell^m(\partialArgr) = \frac{a_\ell^m}{(2\ell - 1)!!} \,\, X_\ell^m(\partialArgr)
\end{equation}
where the constant factors are collected in the quantity
\begin{equation}\label{gloss_2}
  a_\ell^m = (2 - \delta_{m0}) \, \frac{(\ell - \lvert m \rvert)!}{(\ell + \lvert m \rvert)!}
\end{equation}

The Legendre polynomials $\mathcal{P}_\ell(\cos\gamma)$ and the unnormalized real spherical harmonics are related through the equation\cite{orange_book_pisani, math_phys_arfken}
\begin{equation}\label{gloss_1}
  \mathcal{P}_\ell(\cos\gamma) = \sum_{m = -\ell}^{\ell} a_\ell^m \,\, X_\ell^m(\theta, \varphi) \, X_\ell^m(\theta_s, \varphi_s) 
\end{equation}
where $(\theta, \varphi)$ and $(\theta_s, \varphi_s)$ denote two different directions in the spherical coordinate system, separated by an angle $\gamma$, while the expansion coefficients are given by Eq.\nosep\eqref{gloss_2}.

\subsection{Special functions}

The Gamma function is defined following the definition in equation (6.1.1) of Ref.\nosep\cite{abramowitz_stegun_1964}, that is
\begin{equation}\label{gloss_9}
	\Gamma(a) = \int_0^\infty du \,\, u^{a-1} \, e^{-u}
\end{equation}

The error function is defined as in equation (7.1.1) of Ref.\nosep\cite{abramowitz_stegun_1964},
\begin{equation}\label{gloss_10}
	\text{erf}(z) = \frac{2}{\sqrt{\pi}} \int_0^z dt \,\, e^{-t^2}
\end{equation}
while the complementary error function is given by the definition (7.1.2) of Ref.\nosep\cite{abramowitz_stegun_1964}, namely
\begin{equation}\label{gloss_11}
	\text{erfc}(z) = 1 - \text{erf}(z)
\end{equation}

\subsection{Charge and dipole moment}

The charge $N_{\Omegasmall}[n]$ enclosed in the region of space $\Omega$ is given by
\begin{equation}\label{gloss_7}
	\charge_{\Omegasmall}[n] = \int_{\Omega} d^3r \,\, n(\bm{r}) 
\end{equation}
while the total charge $N[n]$ is defined as
\begin{equation}\label{gloss_7.1}
	\charge[n] = \int_{-\infty}^\infty d^3r \,\, n(\bm{r}) 
\end{equation}
Analogously, the dipole moment $\dipole_{\Omegasmall}[n]$ associated to the region of space $\Omega$ is expressed as
\begin{equation}\label{gloss_8}
	\dipole_{\Omegasmall}[n] = \int_{\Omega} d^3r \,\, n(\bm{r}) \, \bm{r}
\end{equation}
and likewise for the total dipole moment  $\dipole[n]$.

\section{Neumann-Laplace expansion of the inverse distance}\label{sec:neumann_laplace_derivation}

The inverse distance between two points in space, appearing on the right hand side of Eq.\nosep\eqref{ele_long_1}, can be expressed in an exact way using the Neumann-Laplace expansion (see equation ($12.4a$) at page 639 of Ref.\nosep\cite{math_phys_arfken}) as follows
\begin{equation}\label{ele_long_2}
	\begin{aligned}
		\frac{1}{\lVert \bm{r} - \bm{r}' - \bm{g} \rVert}
		& = \frac{1}{r_>} \, \sum_{\ell = 0}^\infty \left( \frac{r_<}{r_>} \right)^{\ell} \mathcal{P}_{\ell}(\cos\gamma)
	\end{aligned}
\end{equation}
with $r_< = \min(\lVert \bm{r} - \bm{g} \rVert, \lVert \bm{r}' \rVert)$ and $r_> = \max(\lVert \bm{r} - \bm{g} \rVert, \lVert \bm{r}' \rVert)$, while $\mathcal{P}_{\ell}(\cos\gamma)$ are the Legendre polynomials, where $\gamma$ represents the angle between the vectors $(\bm{r} - \bm{g})$ and $\bm{r}'$ that define the distance in Eq.\nosep\eqref{ele_long_2}.\cite{math_phys_arfken} 
In the case of electrostatic potential at long range, the charge density $n(\bm{r}')$ in Eq.\nosep\eqref{ele_long_1} is localized in a region of space $\Omega$ that is far from the point $(\bm{r} - \bm{g})$ at which the electrostatic potential $\Phi[n](\bm{r} - \bm{g})$ has to be evaluated. Therefore, in the case of electrostatic potential at long range, the inverse distance in Eq.\nosep\eqref{ele_long_1} can be written, using the Neumann-Laplace expansion of Eq.\nosep\eqref{ele_long_2}, as
\begin{align}\label{ele_long_3}
		& \text{Case $\lVert \bm{r} - \bm{g} \rVert \gg \lVert \bm{r}' \rVert$ :} \\
		& \frac{1}{\lVert \bm{r} - \bm{r}' - \bm{g} \rVert} = \frac{1}{\lVert \bm{r} - \bm{g} \rVert} \, \sum_{\ell = 0}^\infty \left( \frac{\lVert \bm{r}' \rVert}{\lVert \bm{r} - \bm{g} \rVert} \right)^{\ell} \mathcal{P}_{\ell}(\cos\gamma) \notag
\end{align}

Inserting the form in Eq.\nosep\eqref{gloss_1} for the Legendre polynomial inside Eq.\nosep\eqref{ele_long_3}, and then using Eq.\nosep\eqref{gloss_4} for the unnormalized real solid spherical harmonics, Eq.\nosep\eqref{ele_long_3} can be rewritten as
\begin{align}\label{ele_long_4}
		& \frac{1}{\lVert \bm{r} - \bm{r}' - \bm{g} \rVert} \\
		& \quad = \sum_{\ell = 0}^\infty \, \lVert \bm{r} - \bm{g} \rVert^{-(2\ell + 1)} \sum_{m = -\ell}^{\ell} a_\ell^m \,\, X_\ell^m(\bm{r} - \bm{g}) \, X_\ell^m(\bm{r}') \notag
\end{align}

Eq.\nosep\eqref{ele_long_4} can be further manipulated, noticing that the inverse distance function is invariant under the translation of both coordinates $\bm{r}$ and $\bm{r}'$ by a constant position vector (that in the following will be indicated as $\centerpos$).
Furthermore, using the Hobson theorem (see Sections 79 and 80 in Ref.\nosep\cite{hobson_1931}), whose application leads to the useful relation
\begin{equation}
\begin{aligned}
	& X_\ell^m(\partialArgRc) \, \lVert \bm{r} - \mathbf{R} \rVert^{-1} \\
	& \qquad \quad = (2\ell - 1)!! \, \lVert \bm{r} - \mathbf{R} \rVert ^{- (2\ell + 1)} \,\, X_{\ell}^m(\bm{r} - \mathbf{R})
\end{aligned}
\end{equation} 
where $X_\ell^m(\partialArgRc)$ is the real spherical gradient operator as defined in Eq.\nosep\eqref{gloss_5}, the product of the powers of the distance function times the first unnormalized real solid spherical harmonics in the expansion of Eq.\nosep\eqref{ele_long_4} can be rewritten as
\begin{equation}\label{ele_long_5}
	\begin{aligned}
		& \lVert \bm{r} - \centerpos - \bm{g} \rVert^{-(2\ell + 1)} \, X_\ell^m(\bm{r} - \centerpos - \bm{g}) \\
		& \qquad = \frac{1}{(2\ell - 1)!!} \,\, X_\ell^m(\partialArgRc) \, \lVert \bm{r} - \centerpos - \bm{g} \rVert^{-1} \\
		& \qquad = \left( a_\ell^m \right)^{-1} \,\, Z_\ell^m(\partialArgRc) \, \lVert \bm{r} - \centerpos - \bm{g} \rVert^{-1}
	\end{aligned}
\end{equation}
where $Z_\ell^m(\partialArgRc)$ is the normalized real spherical gradient operator, as defined in Eq.\nosep\eqref{gloss_6}. Substituting Eq.\nosep\eqref{ele_long_5} inside Eq.\nosep\eqref{ele_long_4}, the inverse distance function of Eq.\nosep\eqref{ele_long_4} can be further rewritten as
\begin{align}\label{ele_long_6}
\begin{aligned}
	& \frac{1}{\lVert \bm{r} - \bm{r}' - \bm{g} \rVert} = \frac{1}{\lVert (\bm{r} - \bm{g} - \centerpos) - (\bm{r}' - \centerpos) \rVert} \\
	& = \sum_{\ell = 0}^\infty \sum_{m = -\ell}^{\ell} X_\ell^m(\bm{r}' - \centerpos) \, Z_\ell^m(\partialArgRc) \, \lVert \bm{r} - \centerpos - \bm{g} \rVert^{-1} \,\,\,
\end{aligned}
\end{align}
thus leading to Eqs.\nosep\eqref{ele_long_7} and \eqref{ele_long_8} of the main text.

\section{Derivation of the Ewald electrostatic potential}\label{sec:ewald_derivation}

The presence of the exponential convergence factor in Eq.\nosep\eqref{ewald_1} leads to the possibility to exchange the summation with the integral sign, thus leading to the equation
\begin{equation}\label{ewald_24}
	\Phi[n](\bm{r},s) = \int_{\Omega} \, d^3r' \,\, n(\bm{r}') \, \psi(\bm{r} - \bm{r}', s)
\end{equation}
where the term $\psi(\bm{x}, s)$ contains the direct space summation and it is defined as
\begin{equation}\label{ewald_25}
	\psi(\bm{x}, s) = \sum_{\bm{g}} \frac{e^{-s\lVert \bm{g} \rVert^2}}{\lVert \bm{x} - \bm{g} \rVert}
\end{equation}

The form of the electrostatic potential in Eq.\nosep\eqref{ewald_0} can be thus recovered from Eq.\nosep\eqref{ewald_24} taking the limit
\begin{equation}\label{ewald_26}
\begin{aligned}
	\Phi[n](\bm{r}) & = \lim_{s \to 0} \Phi[n](\bm{r},s) \\
	& = \int_{\Omega} \, d^3r' \,\, n(\bm{r}') \, \lim_{s \to 0} \psi(\bm{r} - \bm{r}', s)
\end{aligned}
\end{equation}

In the following, the quantity $\psi(\bm{x}, s)$ defined in Eq.\nosep\eqref{ewald_25} will be rewritten, so as to conveniently take the limit of Eq.\nosep\eqref{ewald_26} eliminating, at the same time, the divergence of the original lattice series discussed in Section \ref{sec:integral_test}.
First of all, using the definition of the Gamma function, given by Eq.\nosep\eqref{gloss_9}, with the substitution $u = ty^2$ (being $y$ a constant parameter), the following relation can be easily demonstrated
\begin{equation}\label{ewald_4}
	y^{-2a} = \frac{1}{\Gamma(a)} \int_0^\infty dt \,\, t^{a-1} \, e^{-ty^2}
\end{equation}

\vspace{0.2cm}

Using Eq.\nosep\eqref{ewald_4} to rewrite the inverse distance function in Eq.\nosep\eqref{ewald_25} with $y = \lVert \bm{x} - \bm{g} \rVert$ and $a = 1/2$ gives
\begin{equation}\label{ewald_5}
	\frac{1}{\lVert \bm{x} - \bm{g} \rVert} = \frac{1}{\sqrt{\pi}} \int_0^\infty dt \,\, t^{-1/2} \, e^{-t\lVert \bm{x} \, - \, \bm{g} \rVert^2}
\end{equation}
where  $\Gamma(1/2) = \sqrt{\pi}$ has been used.
Substituting Eq.\nosep\eqref{ewald_5} for the inverse distance in Eq.\nosep\eqref{ewald_25}, it becomes
\begin{equation}\label{ewald_6}
	\psi(\bm{x},s) = \frac{1}{\sqrt{\pi}} \sum_{\bm{g}} {e^{-s\lVert \bm{g} \rVert^2}} \int_0^\infty dt \,\, t^{-1/2} \, e^{-t\lVert \bm{x} \, - \, \bm{g} \rVert^2}
\end{equation}

The integral in Eq.\nosep\eqref{ewald_6} is singular for $s = 0$, at the $t = 0$ limit of the integral. To further isolate the singularity, the integration range in Eq.\nosep\eqref{ewald_6} can be separated in two ranges $[0,\kappa]$ and $[\kappa,\infty)$, respectively defined in the two terms on the right hand side of the following expression
\begin{equation}\label{ewald_7}
	\psi(\bm{x},s,\kappa) = \psi_d(\bm{x},s,\kappa) + \psi_u(\bm{x},s,\kappa)
\end{equation}

The second term on the right hand side of Eq.\nosep\eqref{ewald_7}, related to the integration domain $[\kappa, \infty)$, can be written in terms of the complementary error function of Eq.\nosep\eqref{gloss_11}, using the substitution $v = \sqrt{t} \, \lVert \bm{x} - \bm{g} \rVert$ as follows
\begin{align}\label{ewald_10}
		& \psi_u(\bm{x},s,\kappa)
		= \frac{1}{\sqrt{\pi}} \sum_{\bm{g}} {e^{-s\lVert \bm{g} \rVert^2}} \int_{\kappa}^\infty dt \,\, t^{-1/2} \, e^{-t\lVert \bm{x} \, - \, \bm{g} \rVert^2} \notag \\
		& \qquad = \frac{2}{\sqrt{\pi}} \sum_{\bm{g}} \frac{e^{-s\lVert \bm{g} \rVert^2}}{\lVert \bm{x} - \bm{g} \rVert} \int_{\sqrt{\kappa}\,\lVert \bm{x} \, - \, \bm{g} \rVert}^\infty dv \,\, e^{-v^2} \\
		& \qquad = \sum_{\bm{g}} \, \frac{\text{erfc}\left(\sqrt{\kappa}\,\lVert \bm{x} - \bm{g} \rVert\right)}{\lVert \bm{x} - \bm{g} \rVert} \,\, e^{-s\lVert \bm{g} \rVert^2} \notag
\end{align}

The asymptotic expansion of the complementary error function for large values of the argument $z$ can be evaluated as in equation (7.1.23) in Ref.\nosep\cite{abramowitz_stegun_1964}, thus ensuring that the lattice summation in Eq.\nosep\eqref{ewald_10} is absolutely and uniformly convergent on $s \ge 0$ for $\kappa > 0$. Hence, the limit $s \to 0$ of the second term on the right hand side of Eq.\nosep\eqref{ewald_7} can be easily taken as
\begin{equation}\label{ewald_11}
	\begin{aligned}
		\psi_u(\bm{x},\kappa) & = \lim_{s \to 0} \psi_u(\bm{x},s,\kappa) \\
		& \qquad = \sum_{\bm{g}} \, \frac{\text{erfc}\left(\sqrt{\kappa} \, \lVert \bm{x} - \bm{g} \rVert\right)}{\lVert \bm{x} - \bm{g} \rVert}
	\end{aligned}
\end{equation}
At this point, the singularity at $s = 0$ remains in the first term $\psi_d(\bm{x},s,\kappa)$ on the right hand side of Eq.\nosep\eqref{ewald_7}, related to the integration domain $[0, \kappa]$ of the integral in Eq.\nosep\eqref{ewald_6}. In this case, the dependence on the direct lattice vector $\bm{g}$ of the two exponential factors in Eq.\nosep\eqref{ewald_6} can be collected in a single exponential term by means of the identity
\begin{equation}\label{ewald_12}
\begin{aligned}
	& t\lVert \bm{x} - \bm{g} \rVert^2 + s\lVert \bm{g} \rVert^2 \\
	& \qquad \quad = (t + s)\left\lVert\frac{t\,\bm{x}}{t+s} - \bm{g} \, \right\rVert^2  + \frac{st\lVert \bm{x}\rVert^2}{(t+s)}
\end{aligned}
\end{equation}
thus leading to the following expression
\begin{equation}\label{ewald_8}
	\begin{aligned}
		& \psi_d(\bm{x},s,\kappa) \\
		& \quad \,\, = \frac{1}{\sqrt{\pi}} \sum_{\bm{g}} {e^{-s\lVert \bm{g} \rVert^2}} \int_0^{\kappa} dt \,\, t^{-1/2} \, e^{-t\lVert \bm{x} \, - \, \bm{g} \rVert^2} \\
		& \quad \stackrel{\text{\tiny\eqref{ewald_12}}}{=} \frac{1}{\sqrt{\pi}} \int_0^{\kappa} dt \,\, \Theta(\bm{x}, t, s) \, t^{-1/2} \, e^{-st\lVert \bm{x} \rVert^2/(t\,+\,s)}
	\end{aligned}
\end{equation}
where $\Theta(\bm{x}, t, s)$ contains the direct lattice vectors summation,
\begin{align}\label{ewald_14_1}
	\Theta(\bm{x}, t, s) & = \sum_{\bm{g}} e^{-(t\,+\,s)\lVert t\bm{x}/(t\,+\,s) \, - \, \bm{g} \rVert^2}
\end{align}
Letting $\bm{p} = t\bm{x}/(t + s)$ as well as $\lambda = (t + s)$, and expanding \eqref{ewald_14_1} in a Fourier series (see Section 2.3 of Ref.\nosep\cite{ziman_1972}), the following relation can be obtained
\begin{equation}\label{theta_transformation}
    \sum_{\bm{g}}  e^{-\lambda\lVert \bm{p} \, - \, \bm{g} \rVert^2} = \frac{{\pi}^{3/2}}{v \, \lambda^{3/2}} \sum_{\mathbf{G}} e^{-\lVert \mathbf{G} \rVert^2/(4\lambda)} \,\, e^{i\mathbf{G}\cdot\bm{p}}
\end{equation}
where $\mathbf{G}$ are the reciprocal space vectors and $v$ is the volume of the unit cell. Using the relation \eqref{theta_transformation}, equation \eqref{ewald_14_1} can be rewritten as
\begin{align}\label{ewald_14}
		\Theta(\bm{x}, t, s) & = \sum_{\bm{g}} e^{-(t\,+\,s)\lVert t\bm{x}/(t\,+\,s) \, - \, \bm{g} \rVert^2} \\
		& \stackrel{\text{\tiny\eqref{theta_transformation}}}{=} \frac{\pi^{3/2}}{v (t+s)^{3/2}} \sum_{\mathbf{G}} e^{-\lVert\mathbf{G}\rVert^2/[4(t\,+\,s)]}\,\,e^{it\mathbf{G}\cdot\bm{x}/(t\,+\,s)} \notag
\end{align}

Inserting Eq.\nosep\eqref{ewald_14} in Eq.\nosep\eqref{ewald_8}, the integral results singular for $\mathbf{G} = \bm{0}$ and $s = 0$ at small $t$ values. Therefore, it is convenient to separate the singular term by rewriting the reciprocal lattice summation in Eq.\nosep\eqref{ewald_14} as the sum of the $\mathbf{G} = \bm{0}$ term plus the other non singular terms, so that Eq.\nosep\eqref{ewald_8} can be divided into two terms
\begin{equation}\label{ewald_15}
\begin{aligned}
	\psi_d(\bm{x}, \kappa) & = \lim_{s \to 0} \psi_d(\bm{x}, s, \kappa) \\
	& = \psi_{d,a}(\bm{x}, \kappa) + \psi_{d,b}(\bm{x}, \kappa)
\end{aligned}
\end{equation}
where the first term on the right hand side is obtained excluding the singular term $\mathbf{G} = \bm{0}$ in the reciprocal lattice summation of Eq.\nosep\eqref{ewald_14} to be inserted in the integral of Eq.\nosep\eqref{ewald_8}, and then taking the limit for $s \to 0$, thus leading to the result
\begin{equation}\label{ewald_16}
	\begin{aligned}
		\psi_{d,a}(\bm{x},\kappa) & = \frac{\pi}{v} \, \sum_{\mathbf{G} \neq \bm{0}} \, e^{i\mathbf{G}\cdot\bm{x}} \int_0^{\kappa} dt \,\, t^{-2} \, e^{-\lVert\mathbf{G}\rVert^2/(4t)} \\
		& = \frac{4\pi}{v} \, \sum_{\mathbf{G} \neq \bm{0}} \left( \frac{e^{-\lVert\mathbf{G}\rVert^2/(4\kappa)}}{\lVert \mathbf{G} \rVert^2} \right) e^{i\mathbf{G}\cdot\bm{x}}
	\end{aligned}
\end{equation}
where the integral remained in the first line of Eq.\nosep\eqref{ewald_16} has been easily evaluated analytically, leading to an exponentially convergent series, as reported in the second line of Eq.\nosep\eqref{ewald_16}. The singularity is now contained in the second term on the right hand side of Eq.\nosep\eqref{ewald_15}, whose form is obtained considering only the term $\mathbf{G} = \bm{0}$ in the reciprocal series of Eq.\nosep\eqref{ewald_14}, and then inserting the resultant expression in the last integral of Eq.\nosep\eqref{ewald_8}, thus obtaining
\begin{align}\label{ewald_17}
		& \psi_{d,b}(\bm{x},\kappa) \\
		& \,\, = \frac{\pi}{v} \, \lim_{s \to 0} \int_0^{\kappa} dt \,\, t^{-1/2} \,\, (t+s)^{-3/2} \,\, e^{-st\lVert\bm{x}\rVert^2/(t\,+\,s)} \notag \\
		& \,\, = \frac{\pi}{v} \, \lim_{s \to 0} \left[ \frac{1}{s} \left( \frac{\sqrt{\pi}}{\lVert \bm{x} \rVert \sqrt{s}} \right) \text{erf}\left( \lVert \bm{x} \rVert \sqrt{s} \,\, \sqrt{\frac{\kappa}{\kappa + s}} \, \right) \right] \notag
\end{align}
where the substitution $v = \lVert\bm{x}\rVert [st/(t\,+\,s)]^{1/2}$ together with Eq.\nosep\eqref{gloss_10} have been used to rewrite the integral in Eq.\nosep\eqref{ewald_17} as the error function. By expanding the term in the square parenthesis in the last expression of Eq.\nosep\eqref{ewald_17}, using the Maclaurin series for the error function up to the second term (see equation (7.1.5) in Ref.\nosep\cite{abramowitz_stegun_1964}), and then employing the power series (1.110) of Ref.\nosep\cite{grad_einstein} up to the first order, gives
\begin{align}\label{ewald_18}
		& \frac{1}{s} \left( \frac{\sqrt{\pi}}{\lVert \bm{x} \rVert \sqrt{s}} \right) \text{erf}\left( \lVert \bm{x} \rVert \sqrt{s} \,\, \sqrt{\frac{\kappa}{\kappa + s}} \, \right) \notag \\
		& \quad = \frac{2}{s} \,\, \left( \frac{\kappa}{\kappa + s} \right)^{1/2} - \frac{2 \lVert \bm{x} \rVert^2}{3} \left(\frac{\kappa}{\kappa + s}\right)^{3/2} + O(s) \notag \\
		& \quad = \frac{2}{s} - \frac{1}{\kappa} - \frac{2 \lVert \bm{x} \rVert^2}{3} + O(s)
\end{align}
Substituting this result in the last expression of Eq.\nosep\eqref{ewald_17} leads to the form
\begin{equation}\label{ewald_19}
	\psi_{d,b}(\bm{x},\kappa) = - \frac{\pi}{v \kappa} - \frac{2\pi}{3v}\lVert \bm{x} \rVert^2 + \lim_{s \to 0} \left( \frac{2\pi}{v s} \right)
\end{equation}
Collecting the results of Eqs.\nosep\eqref{ewald_11}, \eqref{ewald_15}, \eqref{ewald_16} and \eqref{ewald_19} inside Eq.\nosep\eqref{ewald_7}, the limit for $s \to 0$ of the quantity in Eq.\nosep\eqref{ewald_7}, to be inserted in Eq.\nosep\eqref{ewald_26} for the electrostatic potential, can be finally written, taking $\bm{x} = \bm{r} - \bm{r}'$, as
\begin{align}\label{ewald_20}
		& \lim_{s \to 0} \psi(\bm{r} - \bm{r}',s,\kappa) \\
		& = \Xi(\bm{r} - \bm{r}', \kappa) - \frac{\pi}{v \kappa} - \frac{2\pi}{3v}\lVert \bm{r} - \bm{r}' \rVert^2 + \lim_{s \to 0} \left( \frac{2\pi}{v s} \right) \notag
\end{align}
where 
\begin{equation}\label{ewald_21}
	\begin{aligned}
		\Xi(\bm{r} - \bm{r}', \kappa) & = \sum_{\bm{g}} \frac{\text{erfc}\left(\sqrt{\kappa} \, \lVert \bm{r} - \bm{r}' - \bm{g} \rVert\right)}{\lVert \bm{r} - \bm{r}' - \bm{g} \rVert} \\
		& + \frac{4\pi}{v} \, \sum_{\mathbf{G} \neq \bm{0}} \left( \frac{e^{-\lVert\mathbf{G}\rVert^2/(4\kappa)}}{\lVert \mathbf{G} \rVert^2} \right) e^{i\mathbf{G}\cdot(\bm{r} - \bm{r}')}
	\end{aligned}
\end{equation}

Finally, the resultant limit of Eq.\nosep\eqref{ewald_20} can be inserted in Eq.\nosep\eqref{ewald_26} for the electrostatic potential, leading to
\begin{align}\label{ewald_27}
	\Phi^{\text{ew}}[n](\bm{r}) & = \int_{\Omega} d^3r' \,\, n(\bm{r}') \,\, \Xi(\bm{r} - \bm{r}',\kappa) \\
	& - \frac{2\pi}{3v} \int_{\Omega} d^3r' \,\, n(\bm{r}') \, \lVert \bm{r}' \Vert^2 \, + \, \frac{4\pi}{3v} \, \bm{r} \cdot \dipole_{\Omegasmall}[n] \quad \notag \\
	& - \left[ \, \frac{2\pi}{3v} \lVert \bm{r} \rVert^2 \, + \, \frac{\pi}{v \kappa} \, - \, \lim_{s \to 0} \left(\frac{2\pi}{v s}\right) \right] \charge_{\Omegasmall}[n] \notag
\end{align}
where Eqs.\nosep\eqref{gloss_7} and \eqref{gloss_8} have been used, respectively for the unit cell charge $\charge_{\Omegasmall}[n]$ and the unit cell dipole moment $\dipole_{\Omegasmall}[n]$.
Using the hypotheses of charge neutrality and zero dipole moment of the unit cell, namely 
\begin{equation}\label{ewald_28}
	\charge_{\Omegasmall}[n] = \mathcal{D}_{i \Omegasmall}[n] = 0 \qquad \qquad \forall \,\, i
\end{equation}
the second term in the second line on the right hand side of Eq.\nosep\eqref{ewald_27}, as well as all the terms in the third line of Eq.\nosep\eqref{ewald_27}, vanish, thus leading to
\begin{equation}\label{ewald_23}
	\begin{aligned}
		\Phi^{\text{ew}}[n](\bm{r}) \,\, & \stackrel{\text{\tiny\eqref{ewald_28}}}{=} \,\, \int_{\Omega} d^3r' \,\, n(\bm{r}') \,\, A(\bm{r} - \bm{r}',\kappa) \\
		& \qquad - \frac{2\pi}{3v} \, \int_{\Omega} d^3r' \,\, n(\bm{r}') \, \lVert \bm{r}' \Vert^2
	\end{aligned}
\end{equation}
where 
\begin{equation}\label{ewald_29}
	A(\bm{r} - \bm{r}',\kappa) \, = \,\, \Xi(\bm{r} - \bm{r}',\kappa) - \frac{\pi}{v \kappa} 
\end{equation}

The resultant Eqs.\nosep\eqref{ewald_23} and \eqref{ewald_29} correspond to Eqs.\nosep\eqref{ewald_9}, \eqref{cell_spheropole} and \eqref{ewald_13} reported in the main text of the article.

\section{Analytical solution of the angular integral \eqref{nmodel_10}}\label{sec:angular_integral}

The angular integral
\begin{equation}\label{ang_int_1}
    \mathcal{A}_{\ell \ell'}^{m m'} = \int_0^{2\pi} d\varphi \int_0^\pi d\theta \,\, X_\ell^m(\theta, \varphi) \, X_{\ell'}^{m'}(\theta, \varphi) \, \sin\theta
\end{equation}
can be solved using the relation \eqref{gloss_3} between the unnormalized real spherical harmonics $X_\ell^m(\theta, \varphi)$ and the unnormalized associated Legendre polynomials $\mathcal{P}_\ell^{m}(\cos\theta)$, so that it can be separated into two independent integrals for the $\theta$ and the $\varphi$ variables, that is
\begin{equation}\label{ang_int_2}
    \mathcal{A}_{\ell \ell'}^{m m'} = \mathcal{I}_{mm'} \, \int_0^\pi d\theta \,\, \mathcal{P}_\ell^{\lvert m \rvert}(\cos\theta) \, \mathcal{P}_{\ell'}^{\lvert m' \rvert}(\cos\theta) \, \sin\theta
\end{equation}
where the first integral in Eq.\nosep\eqref{ang_int_2} is related to the $\varphi$ variable as
\begin{equation}\label{ang_int_3}
	\mathcal{I}_{mm'} = \int_0^{2\pi} d\varphi \,\, t_{m}(\varphi) \, t_{m'}(\varphi) = \frac{2\pi}{\left( 2 - \delta_{m0} \right)} \,\, \delta_{mm'}
\end{equation}
Finally, applying the Kronecker delta coming from the integral \eqref{ang_int_3} inside the remaining integral in Eq.\nosep\eqref{ang_int_2}, together with the Theorem 3.11 of Ref.\nosep\cite{bell_1968}, that is
\begin{equation}\label{ang_int_4}
    \int_{-1}^1 dx \,\, \mathcal{P}_\ell^{\lvert m \rvert}(x) \, \mathcal{P}_{\ell'}^{\lvert m \rvert}(x) = \frac{2(\ell + \lvert m \rvert)!}{(2\ell + 1)(\ell - \lvert m \rvert)!} \,\, \delta_{\ell\ell'}
\end{equation}
the remaining integral in Eq.\nosep\eqref{ang_int_2} can be simply solved using the change of variable $x = \cos\theta$, thus leading to the expression
\begin{equation}\label{ang_int_5}
    \mathcal{A}_{\ell \ell'}^{m m'} = \frac{2(\ell + \lvert m \rvert)!}{(2\ell + 1)(\ell - \lvert m \rvert)!} \,\, \mathcal{I}_{mm'} \, \delta_{\ell\ell'}
\end{equation}
that can be simply rewritten, inserting the explicit expression of the integral $\mathcal{I}_{mm'}$ as given in Eq.\nosep\eqref{ang_int_3}, as in Eqs.\nosep\eqref{nmodel_10} and \eqref{nmodel_11} of the main article.

\section{Computational timings}\label{sec:computational_timings}

The computational timings to perform the calculations in Table\nosep\ref{tab:results} are reported in Table\nosep\ref{tab:results_timings}, showing a speedup of over three orders of magnitude in getting the converged value of the band gap when using a multipole order $L = 6$ with respect to $L = 2$.

\begin{table*}[!htbp]
    \renewcommand{\arraystretch}{1.4}
    \setlength{\tabcolsep}{10pt}
    \centering
    \begin{tabular}{|lcc|ccccc|}
         \multicolumn{1}{c}{} & \multicolumn{1}{c}{} &
         \multicolumn{1}{c}{} &
         \multicolumn{5}{c}{Time [sec/proc]} \\
         \cline{4-8}
         \multicolumn{1}{c}{Tol.} & $N_{\bm{g}}$ & $N_I$ & $L=2$ & $L=3$ & $L=4$ & $L=5$ & $L=6$ \\
         \hline
         $10^{-6}$ & 81 & 5.314$\,\cdot\,10^5$ & 22 & 22 & 22 & 22 & 22 \\
         $10^{-12}$ & 207 & 8.870$\,\cdot\,10^6$ & 124 & 121 & 120 & 119 & 118 \\
         $10^{-20}$ & 387 & 5.796$\,\cdot\,10^7$ & 752 & 738 & 729 & 723 & 716 \\
         $10^{-24}$ & 477 & 1.085$\,\cdot\,10^8$ & 1604 & 1565 & 1535 & 1533 & 1539 \\
         $10^{-30}$ & 677 & 3.103$\,\cdot\,10^8$ & 3925 & 3795 & 4045 & 3772 & 3783 \\
         $10^{-36}$ & 899 & 7.266$\,\cdot\,10^8$ & 8292 & 8069 & 8064 & 8037 & 7943 \\
         $10^{-42}$ & 1067 & 1.215$\,\cdot\,10^9$ & 15598 & 15726 & 15505 & 15361 & 15002 \\
         $10^{-48}$ & 1301 & 2.202$\,\cdot\,10^9$ & 26676 & 26374 & 26097 & 25783 & 25413 \\
         \hline
    \end{tabular}
    \caption{Same as Table\nosep\ref{tab:results} but for computational timings [in seconds per process]. Each reported value represents the mean computational time over four independent runs. Each run was performed using 16 cores on a single {\footnotesize AMD EPYC 9534 64-C}ore processor node.}\label{tab:results_timings}
\end{table*}

\newpage

%
%


\renewcommand{\bibpreamble}{\textbf{References}\\}

\end{document}